\documentstyle[aps]{revtex}

\begin{document}

\preprint{To be published in Proc.R.Soc.London A, June Issue}

\draft

\title{A Universal Two--Bit Gate for Quantum Computation}

\author{Adriano Barenco}
\address{Clarendon Laboratory, Physics Department, University of
  Oxford, Parks Road, Oxford OX1 3PU, United~Kingdom}

\maketitle

\begin{abstract}
  We prove the existence of a class of two--input, two--output gates any
  one of which is universal for quantum computation.  This is done by
  explicitly constructing the three--bit gate introduced by Deutsch
  [Proc.~R.~Soc.~London.~A {\bf 425}, 73 (1989)] as a network
  consisting of replicas of a single two--bit gate.
\end{abstract}

\pacs{\em To be published in Proc.R.Soc.London A, June 1995}

The fact that quantum mechanical processes allow new types of
computation has been known since 1985~\cite{d85}. In 1992, Deutsch and
Jozsa~\cite{dj92} exhibited a class of problems that can be solved
more rapidly on quantum computers than on classical ones and more
recently Shor~\cite{s94} showed that quantum computers can factor
large composite integers very efficiently, a problem for which no
efficient classical algorithm is known. Quantum algorithms for
factoring threaten the security of public key cryptosystems such as
RSA~\cite{RSA} which are currently considered completely reliable.
This suggests that sooner or later perfect security may only be
obtainable via quantum cryptography~\cite{qcrypt}. Clearly, the
experimental realisation of quantum computation is a most important
issue.

Computational networks built out of quantum--mechanical
gates~\cite{D89} provide a natural framework for constructing quantum
computers. A set of gates is {\em adequate\/} if any quantum
computation ({\em i.e.\/} a unitary operation on an
information--carrying register) can be performed with arbitrary
precision by networks consisting only of replicas of gates from that
set. A gate is {\em universal\/} if by itself it forms an adequate
set, {\em i.e.\/} if any quantum computation can be performed by a
network containing replicas of {\em only\/} this gate.

In classical irreversible computation, there exists a universal
two--input one--output gate (the {\sc nand} gate). In classical
reversible computation~\cite{bennett} there exists a three--bit
universal gate (the Toffoli gate~\cite{toffoli}), but not two--bit
universal gate; moreover there is not even an {\em adequate\/} set of
two--bit gates.  DiVincenzo has shown that a certain set of four gates
each operating on two {\em qubits\/} (two--level quantum systems) is
adequate in quantum computation. Deutsch had already shown that the
operation given in the network's {\em computation basis\/} $\{|000
\rangle,|001 \rangle\ldots |111 \rangle \}$ by the unitary matrix
\begin{equation}
D=
\left(
\begin{array}{cc}
 \mbox{ \hspace{2mm} \Large $\hat{1}$ \hspace{2mm}} &
 \mbox{ \hspace{2mm} \Large $\hat{0}$ \hspace{2mm}} \\
 \mbox{ \hspace{2mm} \Large $\hat{0}$ \hspace{2mm}} &
\begin{array}{cccc}
 1& 0 & 0 &0 \\
 0&1&0&0 \\
 0& 0& i\cos{\theta} & \sin{\theta} \\
 0& 0&  \sin{\theta} &i  \cos{\theta}
\end{array}
\end{array}
\right),
\label{deutschgate}
\end{equation}
with $\theta/\pi$ irrational, defines a three--bit (i.e. three--input,
three--output) gate {\bf D} that is universal.
Here $\hat{1}$ and $\hat{0}$ denote respectively the $4\times4$ unit
matrix and the $4\times 4$ zero matrix. Boldface symbols such as {\bf
  D} denote gates, and plain symbols such as $D$ the unitary
operations performed by the corresponding gates.

In this
paper, we improve on the results of both DiVincenzo and Deutsch by showing
that a single {\em two}--bit gate is universal for quantum computation.
Moreover, we
present a large class of universal two--bit gates, providing evidence
that such gates are very common. This result is relevant both from
theoretical and experimental perspectives. The theoretical analysis of
circuit complexity will be much simplified, and from the experimental
point of view, it establishes that an interaction of one type between two
quantum bits is sufficient to ensure universality.

Consider any two--bit gate {\bf A} whose action is given in the
computation basis $\{|00 \rangle,|01 \rangle,|10 \rangle,|11 \rangle\}$ by
the
unitary matrix
\begin{equation}
   A(\phi,\alpha,\theta)=
\left(
\begin{array}{cccc}
   1 & 0 &0 &0\\
   0 & 1 &0 &0 \\
 0& 0& e^{i \alpha} \cos{\theta} &-i e^{i(\alpha-\phi)} \sin{\theta}
\\
 0& 0& -i e^{i(\alpha+\phi)} \sin{\theta} &e^{i\alpha}  \cos{\theta}
\end{array}
\label{universal}
\right),
\end{equation}
where $\phi$, $\alpha$ and $\theta$ are fixed irrational multiples of
$\pi$ and of each other. We shall show that any such gate is
universal.  The proof is by explicit construction of {\bf D} via the
three--bit gate {\bf V} defined by
\begin{equation}
V(\phi,\alpha,\theta)=
\left(
\begin{array}{cc}
\mbox{ \hspace{2mm} \Large $\hat{1}$ \hspace{2mm}} &
\mbox{ \hspace{2mm} \Large $\hat{0}$ \hspace{2mm}}\\
\mbox{ \hspace{2mm} \Large $\hat{0}$ \hspace{2mm}}&
\begin{array}{cccc}
 1& 0 & 0 &0 \\
 0&1&0&0 \\
 0& 0& e^{i \alpha} \cos{\theta} &-i e^{i(\alpha-\phi)} \sin{\theta}
\\
 0& 0& -i e^{i(\alpha+\phi)} \sin{\theta} &e^{i\alpha}  \cos{\theta}
\end{array}
\end{array}
\right).
\end{equation}

Notice first that when the first qubit (which we call the {\em
  control\/} qubit) is in either of the states $|0 \rangle$ or $|1
\rangle$,
{\bf A} induces no entanglement between the qubits. We can read off
from the form of $A$ that if the control is in the state $|0 \rangle$,
the second qubit (which we refer to as the {\em target\/} qubit) is
unaffected by the gate. And if the control is in the state $|1 \rangle$,
the target undergoes the unitary operation given by the lower diagonal
$2\times2$ block of Eq.~\ref{universal}. This operation is a rotation
of angle $2\theta$ about the axis ${\bf u}=\cos(\phi) {\bf x} +
\sin(\phi) {\bf y}$ of the ``spin'' of the target qubit.

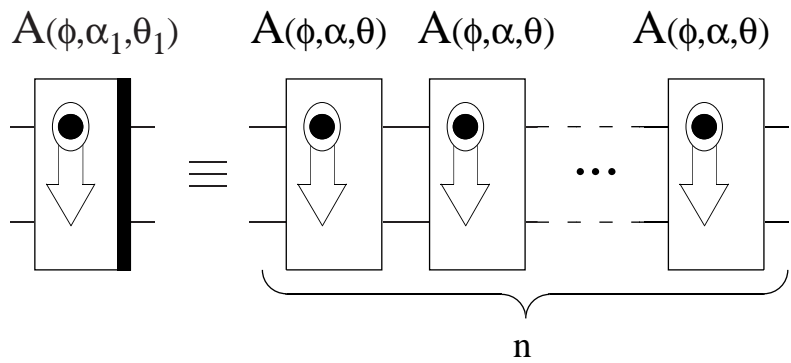
\begin{figure}
\caption[fo0]{Any transformation of the form
  $A(\phi,\alpha_1,\theta_1)$ can be performed with arbitrary accuracy
  by a gate-sequence consisting of $n$ replicas of a single gate {\bf A}
  that effects the unitary transformation $A(\phi,\alpha,\theta)$.
  The control qubits of the gates and gate-sequence are indicated by
  black dots. An arrow points to each target qubit.}
\label{fo0}
\end{figure}

Since
\begin{equation}
A^n(\phi,\alpha,\theta)=
A( \phi,n \alpha \, {\rm mod} \ 2 \pi,n \theta \, {\rm mod}\ 2
\pi),
\end{equation}
and because of the irrationality properties that we required of
$\alpha$ and $\theta$, transformations of the type
$A(\phi,\alpha_1,\theta_1)$, where $\alpha_1$ and $\theta_1$ are any
constants in the range $[0,2\pi]$, can be effected with arbitrary
precision as a result of a sufficient but finite number $n$ of
applications of the operation $A(\phi,\alpha,\theta)$ to the same two
qubits (Fig.~\ref{fo0}).  To do this with $\alpha_1$ and $\theta_1$
specified simultaneously with accuracy $\pm \epsilon$ requires $n \sim
1/\epsilon^2$. The inverse
 $\mbox{\bf A}^{\mbox{{\bf{\scriptsize -1}}}}$
of
  the gate {\bf A}, defined by
\begin{equation}
  A^{-1}(\phi,\alpha,\theta)=A(\phi,2\pi - \alpha,2\pi - \theta),
\end{equation}
is clearly in this repertoire.

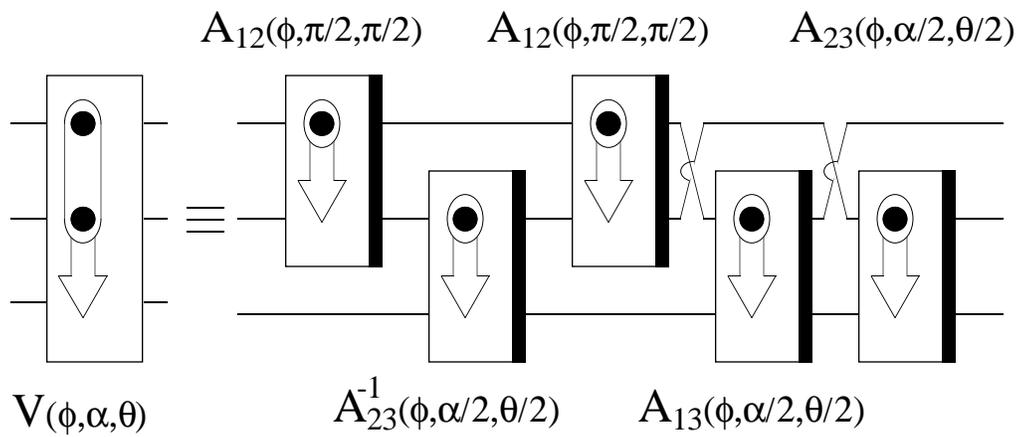
\begin{figure}
\caption[fo1]{A network consisting of sequences of the universal
  two--bit gate {\bf A} has the effect of the three--bit gate
  $V(\phi,\alpha,\theta)$.}
\label{fo1}
\end{figure}

Let $\mbox{\bf A}_{ij}$ denote the three--bit gate obtained from {\bf
  A} by letting qubit $i$ be the control, qubit $j$ the target and
having the remaining qubit go through unaffected. All such gates are
trivially in the repertoire. We have
\begin{equation}
 V(\phi,\alpha,\theta)=
        A_{23}(\phi,\frac{\alpha}{2},\frac{\theta}{2})
        A_{13}(\phi,\frac{\alpha}{2},\frac{\theta}{2})
        A_{12}(\phi,\frac{\pi}{2},\frac{\pi}{2})
        A^{-1}_{23}(\phi,\frac{\alpha}{2},\frac{\theta}{2})
        A_{12}(\phi,\frac{\pi}{2},\frac{\pi}{2}).
\end{equation}
This means that a network of sequences of the gate {\bf A}, as shown
in Fig~\ref{fo1}, has the effect of {\bf V}. This construction, which
greatly simplifies our proof, is similar to that proposed by Sleator
and Weinfurter~\cite{WS94} but uses only one type of gate. {\bf V}
also has a ``control-target'' structure. The state of the third
(``target'') qubit undergoes a non-trivial unitary transformation when
the first two (``control'') qubits are in the state $|11 \rangle$, and
is unaffected if the control qubits are in any of their other three
computation basis states.

If we denote by $\bar{\bf V}$ the gate obtained from {\bf V} by
permuting the second and the third qubit we easily verify that
\begin{equation}
P=\bar{V}(\phi,\pi/2,\pi/2)=
\left(
\begin{array}{cc}
 \mbox{ \hspace{2mm} \Large $\hat{1}$ \hspace{2mm}} &
 \mbox{ \hspace{2mm} \Large $\hat{0}$ \hspace{2mm}} \\
 \mbox{ \hspace{2mm} \Large $\hat{0}$ \hspace{2mm}} &
\begin{array}{cccc}
 1&0&0&0 \\
 0&0&0&e^{-i \phi} \\
 0&0&1&0 \\
 0&e^{i \phi}&0&0
\end{array}
\end{array}
\right)
\end{equation}
and that
\begin{equation}
Q=\bar{V}(\phi,\pi/2,-\pi/2)
V(\phi,\pi/2,-\pi/2)
\bar{V}(\phi,\pi/2,-\pi/2)=
\left(
\begin{array}{cc}
 \mbox{ \hspace{2mm} \Large $\hat{1}$ \hspace{2mm}} &
 \mbox{ \hspace{2mm} \Large $\hat{0}$ \hspace{2mm}} \\
 \mbox{ \hspace{2mm} \Large $\hat{0}$ \hspace{2mm}} &
\begin{array}{cccc}
 1&0&0&0 \\
 0&0&1&0 \\
 0&1&0&0 \\
 0&0&0&1
\end{array}
\end{array}
\right).
\end{equation}

Following the construction of~\cite{D89}, we now note that for
infinitesimal values of $\beta$ the operation
\begin{equation}
  T(\beta)=Q[V(\phi,0,\beta)P ]^2 [V(\phi,0,-\beta)P]^2 Q
\end{equation}
is a rotation of angle $2 \beta^2$ along the axis ${\bf
  u}_\perp=\cos(\phi-\pi/2){\bf x}+\sin(\phi-\pi/2) {\bf y}$ of the
``spin'' of the target qubit :
\begin{equation}
  T(\beta)= 1-i \beta^2 \left(
\begin{array}{cc}
 \mbox{ \hspace{2mm} \Large $\hat{0}$ \hspace{2mm}} &
 \mbox{ \hspace{2mm} \Large $\hat{0}$ \hspace{2mm}} \\
 \mbox{ \hspace{2mm} \Large $\hat{0}$ \hspace{2mm}} &
\begin{array}{cccc}
 0&0&0&0 \\
 0&0&0&0 \\
 0&0&0&i e^{-i \phi} \\
 0&0&-i e^{i \phi}&0
\end{array}
\end{array}
\right)
+ O(\beta^3).
\end{equation}
Therefore the transformation
\begin{equation}
V(\phi-\pi/2,0,\beta)=
  \lim_{n \rightarrow
    \infty}T(\sqrt{\beta/n})^n
\end{equation}
can also be performed with arbitrary accuracy by networks of the gate
{\bf A}. Similarly, so can
\begin{eqnarray}
R_z(\beta)&=&  \lim_{n \rightarrow \infty}
\left[
V(\phi,0,\sqrt{\beta / 2n})
V(\phi-\pi/2,0,\sqrt{\beta / 2n})
V(\phi,0,-\sqrt{\beta / 2n})
V(\phi-\pi/2,0,-\sqrt{\beta / 2n})
\right]^n
\\ \nonumber
&=&
 \left(
\begin{array}{cc}
 \mbox{ \hspace{2mm} \Large $\hat{1}$ \hspace{2mm}} &
 \mbox{ \hspace{2mm} \Large $\hat{0}$ \hspace{2mm}} \\
 \mbox{ \hspace{2mm} \Large $\hat{0}$ \hspace{2mm}} &
\begin{array}{cccc}
 1&0&0&0 \\
 0&1&0&0 \\
 0&0&e^{i \beta}&0 \\
 0&0&0&e^{-i \beta}
\end{array}
\end{array}
\right).
\end{eqnarray}
This last gate is a conditional rotation along the $z$ axis of the
``spin'' of the target qubit when the two controls are in state
$|11 \rangle$.

The universal Deutsch gate {\bf D} may now be constructed:
\begin{equation}
 D=R_z(\phi/2) V(\phi,\pi/2,\theta) R_z(-\phi/2).
\end{equation}

An exceptionally simple two--bit universal quantum gate is the one
that performs the operation $A(\pi/2,\pi/4,\theta)$. Note that this
gate, with $\alpha = \pi/4$, does not even satisfy the irrationality
constraint we imposed, but is nevertheless universal by the same
proof. It is also particularly appealing from the experimental point
of view. The lower diagonal $2\times 2$ block of
$A(\pi/2,\pi/2,\theta)$ is a rotation of angle $\theta/2$ about the
$y$ axis of the ``spin'' of the qubit. In a realistic implementation,
such an operation can be realised by applying a $(\theta/2)$--pulse of
resonant radiation to the spin of the qubit.
A variety of experimentally realisable systems can be put forward as
practical implementations of gates of this type. For instance, one
possibility is to use cavity QED--type of interaction as analysed
recently by Davidovich {\em et al.\/}~\cite{davidovich} in the
context of teleportation, or use ion--ion interactions in linear traps
along a similar scheme as the one recently proposed by Cirac and
Zoller~\cite{CZ95}

However, simplicity in the form of the interaction is not always to
the point. It is desirable that practical gates should have as generic
a form as possible. In this paper, we have presented a
three--parameter family of universal gates. In another
paper~\cite{dbe95}, it will be shown that starting from the result just
presented, further generalisation is
possible: conforming to a conjecture of Deutsch~\cite{D89}, it turns
out that almost all two--bit quantum gates are universal.

Many thanks to D.~Deutsch for numerous discussions and comments at
every stage of this work. Thanks also to D.~DiVincenzo, T.~Sleator and
H.~Weinfurter for sending preprints of their papers and to A.~Ekert
and M.~Palma for reviewing earlier drafts the manuscript.  This work was
supported by
the Berrow's fund at Lincoln College, Oxford, and the Royal Society,
London.

\end{document}